\documentclass{emulateapj}

\slugcomment{Journal Reference: Icarus 252 (2015) 334-338}

\begin{document}

\title{Formation of Phobos and Deimos via a Giant Impact}
\author{Robert I. Citron $^{1}$, Hidenori Genda$^{2}$, and Shigeru Ida$^{2}$}
\affil{$^{1}$Department of Earth and Planetary Science, University of California Berkeley, USA}
\affil{$^{2}$Earth-Life Science Institute, Tokyo Institute of Technology, Japan}


\begin{abstract}
Although the two moons of Mars, Phobos and Deimos, have long been thought to be captured asteroids, recent observations of their compositions and orbits suggest that they may have formed from debris generated by one or more giant impacts of bodies with $\sim$ 0.01 $\times$ target mass. Recent studies have both analytically estimated debris produced by giant impacts on Mars and numerically examined the evolution of circum-Mars debris disks. We perform a numerical study (Smoothed Particle Hydrodynamics simulation) of debris retention from giant impacts onto Mars, particularly in relation to a Borealis-scale giant impact (E $\sim 3 \times 10^{29}$ J) capable of producing the Borealis basin. We find that a Borealis-scale impact is capable of producing a disk of mass $\sim 5 \times 10^{20}$ kg ($\sim$ 1 - 4 \% of the impactor mass), sufficient debris to form at least one of the martian moons according to recent numerical studies of martian debris disk evolution. While a Borealis-scale impact may generate sufficient debris to form both Phobos and Deimos, further studies of the debris disk evolution are necessary. Our results can serve as inputs for future studies of martian debris disk evolution. 
\end{abstract}

\section{Introduction}  \label{introduction}
%

The origin of the two moons of Mars, Phobos and Deimos, is widely debated. One prevailing theory is that the two moons are captured asteroids, as indicated by their size, shape, and prior observations of their composition \citep{1991JGR....96.5925M, 1978sama.conf..193B, 2008pmsa.book.....F}. However, recent observations \citep[e.g.,][]{2011P&SS...59.1308G, 2014P&SS..102...18W} suggest that Phobos and Deimos may have formed from debris generated from re-accretion after one or more giant impacts of bodies with $\sim$ 0.01 $\times$ target mass, a process that could more readily explain the composition, density, and orbits of the martian moons \citep{2011A&ARv..19...44R}. Compositionally, observations from both the European Space Agency's Mars Express mission and NASA's Mars Global Surveyor mission show that Phobos and Deimos have thermal infrared spectra in poor agreement with the primitive carbon-rich materials normally associated with carbonaceous chondrites \citep{2011P&SS...59.1308G, 2014P&SS..102...18W}. Additionally, minerals were detected on Phobos that are also present on the martian surface \citep{2011P&SS...59.1308G, 2014P&SS..102...18W}, suggesting that the material that formed the martian satellites could have been ejected from Mars. Furthermore, the Mars Radio Science Experiment on board Mars Express mission measured the density of Phobos to be $1.87$ g cm$^{-3}$ with a porosity of 25-35\% \citep{Andert:2010p15153, 2014P&SS..102...86P, Rosenblatt:2010p15160} $-$ such low density and high porosity are uncommon for most asteroids. An asteroid with such a composition would not be expected to remain coherent if it was dynamically captured \citep{Craddock:1994p14997}, although it could be expected to form by material re-accreted from a debris disk \citep{Richardson:2002p15201}. Re-accretion could also explain the circularity of the orbits of Phobos and Deimos; if the martian moons were captured objects, they would be expected to have more eccentric orbits and randomly oriented orbital planes -- and although the orbit of Phobos could be circularized via tidal perturbations from Mars, Deimos is too far away to be strongly affected by such perturbations. Additionally, moons that formed from the same impact that produced the current martian spin would be expected to orbit near the equatorial plane, as Phobos and Deimos do \citep{2012Icar..221..806R}. 

The observations discussed above makes a strong case that Phobos and Deimos formed via re-accretion, but this process requires the ejection of sufficient satellite-forming material into circum-Mars orbit, most likely via a giant impact \citep{Craddock:2011p15993}. Giant impacts were a common process in the late stages of planetary formation, and several large basins between 3300 and 4500 km in diameter suggest impacts that could have ejected significant debris into martian orbit. The largest possible impact site is the 7700 km-diameter Borealis basin that encompasses almost the entire northern hemisphere. Although whether the Borealis basin formed from exogenic or endogenic processes is still debated, recent analysis of the martian crust suggests that a giant impact could explain the elliptical shape of the basin and the sharp crustal thickness gradient on the basin perimeter \citep{AndrewsHanna:2008p10071}. In addition, two independent studies numerically simulated Borealis-scale giant impacts (E $\sim 1-3 \times 10^{29}$ J) and found that such an impact could have formed a structure similar to the basin \citep{Marinova:2008p13805, Nimmo:2008p13811}. 

A Borealis-scale impactor would likely have had a mass of $\sim$ 0.026 Mars masses \citep{Marinova:2008p13805}, allowing the ejection of substantial material into martian orbit. The total mass of ejected material necessary to form Phobos and Deimos is unclear, and is complicated by the large number of elongate craters on the surface of Mars that could have formed via impacts by additional short-lived martian satellites \citep{Schultz:1982p15139, Strom:1992p15107}. The total mass of past and present martian satellites, $M_{sats}$, is estimated to be between $8.3 \times 10^{17}$ and $1.5 \times 10^{19}$ kg ($\sim 1 \times 10^{-6} - 2 \times 10^{-5}$ Mars masses), based on the application of the Gault and $\pi$-group crater scaling laws \citep{1974plug.nasa..137G, 1977iecp.symp.1231G, MeloshHJ:1989p38} to various estimates of the number of elongate martian craters attributed to the impact of de-orbited past satellites \citep{Craddock:2011p15993, Schultz:1982p15139, 2000Icar..145..108B}. Although such an estimate is limited because the elongated crater population could also include grazing asteroid impacts and may not account for all prior martian satellites, the estimated mass of past and present satellites is still considerably greater than the current masses of Phobos and Deimos, which are $1.07 \times 10^{16}$ and $1.48 \times 10^{15}$ kg, respectively. 

Recently, the feasibility of forming the martian moons from giant impacts has been studied both analytically \citep{Craddock:2011p15993} and numerically \citep{2012Icar..221..806R}. \citet{Craddock:2011p15993} estimated that the mass of a circum-Mars debris disk should be twice the mass contained in the present and past martian satellites ($M_{sats}/M_d \sim 0.5$), based on the assumption that half of the disk material migrates inward while the other half migrates outward from Mars to form satellites. Comparing this estimate to the impactor mass of $\sim$ 0.02 $M_{Mars}$ that would be necessary to cause the current martian spin-rate, Craddock estimated that $\sim$ 0.01 - 0.4 \% of the impactor mass must be ejected into orbit to form the martian moons, a circum-Mars accretion disk of $\sim 1 \times 10^{18} - 5 \times 10^{19}$ kg ($M_d/M_{imp} \sim 1 \times 10^{-4} - 4 \times 10^{-3}$). 

To determine if such an inserted mass could produce the total estimated mass of prior and current martian satellites, \citet{2012Icar..221..806R} modeled the evolution of a $10^{18}$ kg circum-Mars debris disk, a problem significantly different from prior models of satellite formation in a circum-Earth debris disk \citep[e.g.,][]{Ida:1997p13527} because of the low mass and density of the proposed martian disk. \citet{2012Icar..221..806R} modeled circum-Mars disk evolution in both a strong tidal regime, where accretion occurs at the Roche limit close to the planet, and a weak tidal regime, where accretion occurs further away from the planet. In the strong tidal regime, \citet{2012Icar..221..806R} found that to account for a moonlet population of $8.3 \times 10^{17} - 1.5 \times 10^{19}$ kg, the disk must have had a mass of $8.3 \times 10^{19} -1.5 \times 10^{21}$ kg, $30 - 80$ times the mass predicted by \citet{Craddock:2011p15993}. If the relation, $M_d \sim$ 100 $M_{sats}$, obtained by \citet{2012Icar..221..806R} is correct, in order to be consistent with $M_{imp} \sim$ 0.026 $M_{Mars}$ and $M_{sats} \sim 1 \times 10^{-6} - 2 \times 10^{-5}$, $M_d/M_{imp}$ must be $(M_{sats}/M_{imp}) \times (M_d/M_{sats}) \sim 0.004 - 0.08$. So far, the only other direct calculation of $M_d/M_{imp}$ is by \citet{CanupSalmon:2014}, which was reported at the DPS meeting in 2014. 

To directly determine an estimate for $M_d/M_{imp}$, we performed numerical Smoothed Particle Hydrodynamics (SPH) simulations of giant impacts into Mars. We take our canonical giant impact from \citet{Marinova:2008p13805}, a 45 degree impact with an impact energy $E_{imp} = 3 \times 10^{29}$ J and $M_{imp} = 0.026$ $M_{Mars}$, which has been shown to produce a crustal dichotomy similar to that currently observed on Mars. To test the robustness of the result, we vary impactor angle and impact energy, particularly to account for other proposed mechanisms of dichotomy formation, such as an impact megadome requiring an impact with $E_{imp} \sim 1-3 \times 10^{30}$ J \citep{Reese:2011p16342}. Our numerical study of a range of potential impacts allows for improved inputs into circum-Mars disk evolution models.

\section{SPH simulations} \label{sph}

\subsection{SPH method} \label{sph_method}

We model giant impact scenarios onto Mars using the SPH method \citep{Monaghan:1992p10226}. SPH codes have been used to model several types of giant impacts, including the putative Moon-forming impact \citep{1989Icar...81..113B, 1986Icar...66..515B, 1987Icar...71...30B, 1997Icar..126..126C, Cameron:1991, 2004Icar..168..433C, Canup:2008, 2001Natur.412..708C} and other types of large-scale planetary collisions \citep{Agnor:2004p10045, Genda:2012p18168}. An SPH code has also been used to model the Borealis impact on Mars \citep{Marinova:2008p13805}, but these simulations focused on the structure of the impact basin and did not examine the formation of a debris disk. 

We use the SPH code described in \citet{Genda:2012p18168}, which is based on the convention of \citet{Monaghan:1992p10226}. Our simulations use the Tillotson equation of state (EOS) \citep{Tillotson:1962p16037}, which is widely used in large-scale planetary impact simulations \citep[e.g.,][]{1987Icar...71...30B, Canup:2001p16040, Marinova:2008p13805}. The Tillotson EOS parameters used in our study are identical to those used in prior simulations of a giant impact on Mars \citep{Marinova:2008p13805}. To represent Mars, we use a differentiated iron/mantle target body with a core to mantle mass ratio of 0.3 and a total weight of 6.0 $\times 10^{23}$ kg. The impactor is an undifferentiated basalt body with a mass of 0.026 $\times$ target mass. Our canonical impact has an impact velocity of 6 km/s, corresponding to an impact energy of $3 \times 10^{29}$ J, similar to the nominal simulation from \citet{Marinova:2008p13805}. We vary impact velocity ($\alpha = v_{imp}/v_{esc}$) and impactor mass, $M_{imp}$, to obtain disk masses for several types of giant impacts into Mars.



\begin{deluxetable*}{cccccccccc}
\tablecaption{Disk masses for several SPH simulations with various impact parameters\tablenotemark{a}. }
\tablehead{\colhead{Run} & \colhead{$N$} & \colhead{$M_{imp}$ (kg)} & \colhead{$\alpha$} & \colhead{$\theta$ ($^{\circ}$)} & \colhead{$E_{imp}$ (J)} & \colhead{$\gamma$} & \colhead{$M_d/M_{imp}$  \tablenotemark{b}} & \colhead{$M_d$ (kg) \tablenotemark{b}} & \colhead{$L_d^*$ \tablenotemark{b,c}}}

\startdata
0 & $5\times10^4$ & $1.68\times10^{22}$ & 1.4 & 45 & $3\times10^{29}$ & 0.028 & 0.026 &  $4.3\times10^{20}$ &$0.95$\\
1 & $1\times10^5$ & $1.68\times10^{22}$ & 1.4 & 45 & $3\times10^{29}$& 0.028 & 0.029 &  $4.8\times10^{20}$ &$0.98$\\
2 & $3\times10^5$ & $1.68\times10^{22}$ & 1.4 & 45 & $3\times10^{29}$ & 0.028 & 0.033 &  $5.5\times10^{20}$  &$0.92$\\
3 & $1\times10^6$ & $1.68\times10^{22}$ & 1.4 & 45 & $3\times10^{29}$ & 0.028 & 0.033 &  $5.5\times10^{20}$ &$0.90$\\
4 & $3\times10^5$ & $1.68\times10^{22}$ & 1.4 & 30 &$3\times10^{29}$ & 0.028 & 0.024 &  $4.0\times10^{20}$  &$0.89$\\
5 & $3\times10^5$ & $1.68\times10^{22}$ & 1.4 & 60 & $3\times10^{29}$ & 0.028 & 0.017 &  $2.8\times10^{20}$  &$1.01$\\
6 & $3\times10^5$ & $1.68\times10^{22}$ & 2 & 45 & $6.3\times10^{29}$ & 0.028 & 0.024 &  $4.0\times10^{20}$  &$0.95$\\
7 & $3\times10^5$ & $3.5\times10^{22}$ & 1.4 & 45 & $6\times10^{29}$ & 0.058 & 0.033 &  $1.14\times10^{21}$  &$0.86$\\
8 & $3\times10^5$ & $6.0\times10^{22}$ & 1.4 & 45 & $1\times10^{30}$ & 0.100 & 0.023 &  $1.4\times10^{21}$  &$0.89$\\
9 & $3\times10^5$ & $1.0\times10^{23}$ & 1.9 & 45 & $3\times10^{30}$ & 0.167 & 0.013 &  $1.3\times10^{21}$  &$0.76$\\
\enddata
\tablenotetext{a}{Here, we list our ten simulations according to run number, number of particles in the simulation ($N$), impactor mass ($M_{imp}$), relative impact velocity ($\alpha = v_{imp}/v_{esc}$), impact angle ($\theta$), energy of the impact ($E_{imp}$), impactor-to-target mass ratio ($\gamma = M_{imp}/M_{target}$), disk-to-impactor mass ratio ($M_d/M_{imp}$), disk mass ($M_d$), and normalized angular momentum ($L_d^*$).}
\tablenotetext{b}{$M_d/M_{tot}$, $M_d$, and $L_d^*$ are all reported at $t = 30$ hrs.}
\tablenotetext{c}{We normalize the angular momentum as $L_d^* = L_d/(M_d \sqrt{G M_{Mars} R_{Roche}})$.}
\label{tbl1}
\end{deluxetable*}


After each SPH simulation is complete, we quantify the mass and angular momentum of the impact-generated disk using the method described in \citet{Cuk:2012p18167}; the disk contains SPH particles that have a negative total energy and an equivalent circular orbital radius that is greater than an estimated planetary radius defined by a 1 g cm$^{-3}$ density contour. This method or a similar calculation is generally used to estimate disk masses from large collisions \citep[e.g.,][]{Canup:2008, Cuk:2012p18167}, however, it is only an effective estimate for the initial mass of the disk shortly after impact. For timescales greater than several tens of hours after impact, artificially induced spreading in the SPH simulation leads to an ineffective disk mass calculation.  Furthermore, the SPH simulation is limited in that the collisionless particle representation cannot accurately simulate accretion in a disk, a computation better estimated using numerical models of disk evolution \citep[e.g.,][]{Ida:1997p13527, 2012Icar..221..806R}. Therefore, our initial disk mass estimates are meant to be compared with or used as inputs for long-term disk evolution models, and because the damping of inclination is rapid in comparison to the disk's radial evolution \citep{Ida:1997p13527}, disk mass should not change significantly beyond the time limit of our simulations due to damping. 

\begin{figure}
\includegraphics[width=\linewidth]{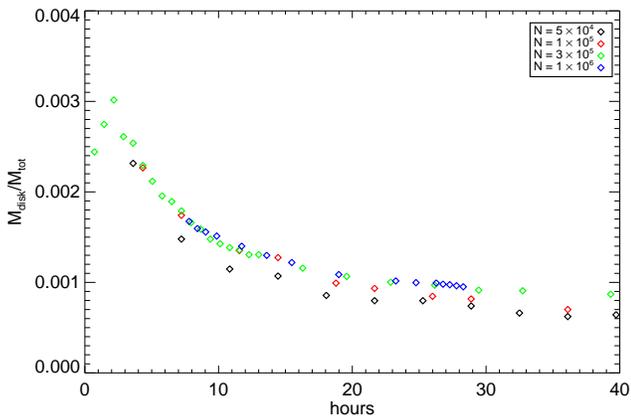}
\caption{Disk mass versus time for several total particle numbers.}
\label{fig:res}
\end{figure}

For our canonical impact, we conducted a resolution test (Runs $0 - 3$, Table \ref{tbl1}) to compare disk mass evolution versus time for several total SPH particle counts (Fig. \ref{fig:res}). For the first 40 hours after the impact, the disk mass evolution of a simulation with $3 \times 10^{5}$ particles is similar to the simulation with $10^{6}$ particles. Because of this, we use a resolution of N = $3 \times 10^5$ particles for the remainder of our simulations.

\subsection{SPH results} \label{sph_results}
Because the proposed Borealis impact is lower energy than the putative Moon-forming collision, impacts studied in this investigation produced much smaller disks than previous studies of Earth-Moon formation. The disk masses for the collision scenarios we investigated are shown in Table \ref{tbl1}. For our canonical impact (Run 2, Table \ref{tbl1}), a disk of mass $5.55 \times 10^{20}$ kg is produced 30 hours post-impact, composed of $\sim 280$ particles. 

\begin{figure}
\centering
\includegraphics[width=\linewidth]{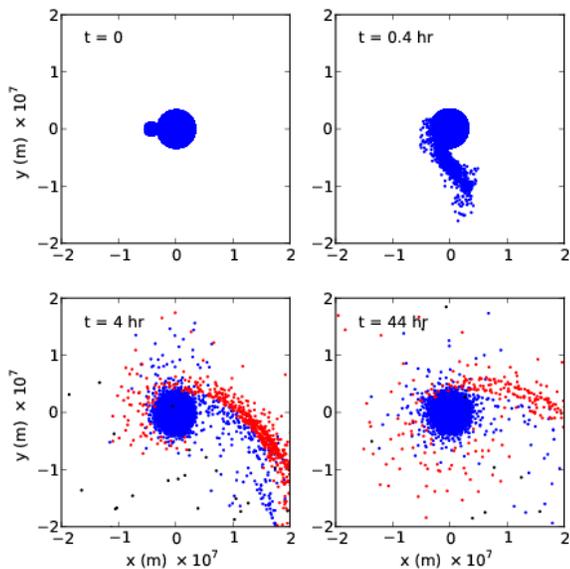}
\caption{Snapshots of an SPH simulation during the impact and disk formation. Ejected particles are shown in black and disk particles are shown in red. The blue particles are considered part of the planet. }
\label{fig:sims}
\end{figure}

\begin{figure}
\centering
\includegraphics[width=\linewidth]{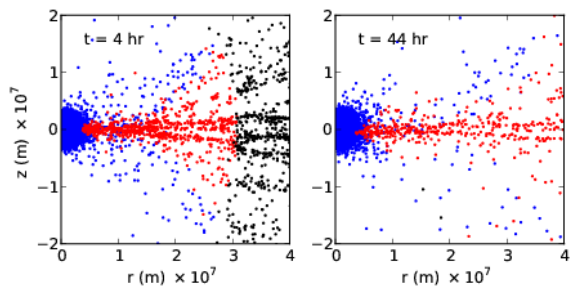}
\caption{An edge on view of the simulation from Fig. 1 at $t$ = 4 and 44 hr. Colors of particles are the same as in Fig. 1.}
\label{fig:simsz}
\end{figure}

An example of the evolution of a debris disk formed after our canonical Borealis-scale impact is shown in Fig. \ref{fig:sims}. In Fig. \ref{fig:sims}, the blue particles represent particles that are part of the planet or lack sufficient angular momentum to be considered part of the disk. The red particles are part of the disk, and black particles have been ejected from the system. The initial shock of the impact creates a large amount of ejecta that orbits the planet. While a degree of spreading of the ejecta occurs, a large amount of ejecta remains relatively close to the planet in a disk. An edge-on view of the disk is also shown in Fig. \ref{fig:simsz}. The disk appears to undergo relatively quick damping, and Fig. \ref{fig:res} shows that the disk mass becomes more constant as simulation time increases. 

To test the robustness of our result, we varied impact angle and impact energy, and compared the resulting debris disk masses. For impact angle, in addition to the nominal 45 degree impact angle, we also tested impacts with angles of 30 and 60 degrees. These changes in impact angle only slightly reduced the disk mass. For impact energy, we ran several simulations with higher impact energy by increasing the impactor mass and/or velocity (Runs 6$-$9, Table \ref{tbl1}). While increasing the impact energy did result in larger mass debris disks, the highest energy impact we examined (Run 9) only resulted in approximately double the disk mass of the canonical impact (Run 2).  In all cases, the amount of disk material produced was about 1 to 4 \% of the impactor mass. Therefore, we can conclude that $M_d/M_{imp} \sim (1-4) \times 10^{-2}$ is robust for Borealis-scale impacts. This appears to agree with the result of \citet{CanupSalmon:2014}, which found that an impactor with a few percent of Mars' mass produced a disk several orders of magnitude more massive than Phobos and Deimos. 

We also examined the disk angular momentum, scaled as $L_d^* = L_d/(M_d \sqrt{G M_{Mars} R_{Roche}})$. If $L_d^*$ is greater than 1, most of the mass is outside of the Roche limit, therefore this quantity represents the compactness of the disk compared to the Roche limit. Our simulations (Table \ref{tbl1}) produce disks with $L_d^* < 1$ for all but one case. We find that $L_d^*$ decreases when both the impactor mass and impact energy increase. The higher angle $\theta = 60^{\circ}$ impact (Run 5) produces the least compact disk. Our canonical simulation (Run 2) produces $L_d^* = 0.92$, representing a slightly compact disk where most of the mass is within the Roche limit, but moderate mass is also present outside of the Roche limit. The presence of mass outside the Roche limit is also reported in the \citet{CanupSalmon:2014} abstract. Because a value of $L_d^* \sim 0.9$ should correspond to the strong tidal regime described by \citet{2012Icar..221..806R}, our results indicate that accretion in the strong tidal regime is favored over accretion in the weak tidal regime. 

\section{Discussion and Conclusion} \label{discussion}

Our examination of a Borealis-scale impact (Run 2, Table \ref{tbl1}) shows that $\sim 5 \times 10^{20}$ kg of disk material is produced 30 hours post-impact, well within the $8.3 \times 10^{19} - 1.5 \times 10^{21}$ kg disk mass \citet{2012Icar..221..806R} require in their strong regime to produce sufficient satellite mass to explain the elongated crater distribution on Mars. Modest adjustments to the impact angle or velocity produce similar disk masses, suggesting that sufficient material to produce Phobos and Deimos, or even the entire past/present satellite population, could have been ejected in a Borealis-scale impact. The larger impacts we examined, analogous to collisions required to form impact megadomes \citep[e.g.,][]{Reese:2011p16342}, produce over twice the disk mass as the smaller scale impacts. This is at the upper limit of the disk mass required to produce the theoretical moonlet population and greater than the disk mass necessary to produce Deimos in the weak accretion regime. 

Our finding that 1 to 4 \% of the impactor mass is inserted into an accretion disk is $20 - 100$ times higher than the \citet{Craddock:2011p15993} analytical estimate that 0.01 to 0.2 \% of the impactor is inserted into orbit. This is partly because \citet{Craddock:2011p15993} estimates the ejected mass by assuming that half of a circum-Mars disk would migrate outward and accrete into satellites, and therefore the disk mass should be twice the mass of the past and present satellites. Additionally, the Gault and $\pi$-group scaling laws used to estimate the mass of past satellites are valid only for 90$^{\circ}$ impacts therefore should only provide a lower limit on the masses of the satellites that could have formed elongate craters on Mars \citep{Craddock:2011p15993}. By numerically modeling the evolution of circum-Mars debris disks, \citet{2012Icar..221..806R} found that only 1\% of the disk mass necessarily accreted into satellites. Therefore, although our simulations produce disks with higher mass than the \citet{Craddock:2011p15993} estimates, if less of the disk mass accretes into moonlets and satellites, then similar satellite mass distributions can be achieved. 

It should be noted that our study examines a narrow parameter space of Borealis-scale impacts that we find produce disks more massive than the $10^{18}$ kg disk examined by \citet{2012Icar..221..806R}. Therefore, although we assume that 1\% of the disk mass forms satellites \citep{2012Icar..221..806R}, it is not certain that this estimate holds at larger disk masses, and further simulations of circum-Mars disk evolution for larger disk masses are necessary. Additionally, accretion in more massive disks ($> 10^{18}$ kg) could produce satellites larger in size than Phobos or Deimos \citep{2012Icar..221..806R}; however, these satellites could eventually de-orbit and impact Mars, depending on where they form in the disk. Furthermore, in addition to Borealis-scale impacts, \citet{Craddock:2011p15993} suggests that the Elysium and Utopia basins could have been formed by large enough impacts to produce the current martian spin rate, suggesting impacts as low as 0.015 Mars masses. A basic power law fit to our simulations with fixed $\alpha = 1.4$ and $\theta = 45$ (Runs 2, 7, and 8) yields $M_d = 8 \times 10^{21} (M_{imp})^{0.75}$, meaning an impactor with 0.015 target masses would produce a disk of mass $\sim 3 \times 10^{20}$ kg. Direct simulations of smaller impacts would be useful, but as the impact energy decreases, more total particles are needed in the simulation if the final disk is to contain at least the couple hundred particles necessary to accurately estimate the disk mass. Because our impact energies fall in the most-likely range reported by \citet{Craddock:2011p15993}, we leave the examination of less energetic impacts for future work.

Our simulations show that for Borealis-scale impacts, enough material is ejected into orbit to form accretion disks that could produce martian satellites. Our result is inconsistent with the formation of only Phobos and Deimos in the weak tidal regime considered by \citet{2012Icar..221..806R} because our disk is $100 - 1000$ times more massive and the $L_d^*$ values of our disks mostly indicate too much compactness for the weak tidal regime. Therefore, our results support accretion in the strong tidal regime. Although additional simulations of disk evolution with disks of initially larger mass are required to further investigate scenarios of martian moon formation, our results provide a numerical estimate of disk mass production in Borealis-scale impacts.

\section{Acknowledgement} \label{acknowledgement}
R. Citron would like to thank the NSF's East Asia and Pacific Summer Institute (Award 1107883) and the Japan Society for the Promotion of Science's Summer Science Program for their support of this research. We thank valuable and helpful comments by anonymous referees.


\bibliographystyle{icarus}
\bibliography{MarsMoonsLib}
\end{document}